\documentclass[aps,twocolumn,floatfix,groupedaddress,nofootinbib]{revtex4}
\usepackage{csquotes}
\usepackage{graphicx,color}
\usepackage{float}
\usepackage[caption=false]{subfig}
\usepackage{amsmath}
\usepackage{amsmath}
\usepackage{multirow}
\usepackage{dsfont}
\usepackage[shortlabels]{enumitem}
\usepackage{epstopdf}

\begin{document}

\title{On the advanced integral equation theory description of dense Yukawa one-component plasma liquids}
\author{F. Lucco Castello and P. Tolias}
\affiliation{Space and Plasma Physics, Royal Institute of Technology, Stockholm, SE-100 44, Sweden\\}
\begin{abstract}
\noindent Different advanced bridge function closures are utilized to investigate the structural and thermodynamic properties of dense Yukawa one-component plasma liquids within the framework of integral equation theory. The isomorph-based empirically modified hypernetted-chain, the variational modified hypernetted-chain, the Rogers-Young and the Ballone-Pastore-Galli-Gazzillo approaches are compared at the level of thermodynamic properties, radial distribution functions and bridge functions. The comparison, based on accuracy and computational speed, concludes that the two modified hypernetted-chain approaches are superior and singles out the isomorph-based variant as the most promising alternative to computer simulations of structural properties of dense Yukawa liquids. The possibility of further improvement through artificial cross-over to exact asymptotic limits is studied.
\end{abstract}
\maketitle

\section{Introduction}\label{sec:intro}

\noindent Yukawa one-component plasmas (YOCP) are model systems consisting of charged point particles that are immersed in a neutralizing background and interact via the pair-wise potential $u(r)=(Q^2/r)\exp(-r/\lambda)$. Here $Q$ is the particle charge and $\lambda$ is the screening length dictated by the polarizable background. It has become customary to specify the thermodynamic YOCP states in terms of two independent dimensionless variables, namely the coupling parameter $\Gamma$ and the screening parameter $\kappa$ that are defined by\,\cite{fortov2005,morfill2009,bonitz2010}
\begin{equation*}
\Gamma=\beta\frac{Q^2}{d}\,,\qquad\kappa=\frac{d}{\lambda}\,,
\end{equation*}
where $\beta=1/(k_{\mathrm{B}}T)$ with $T$ the temperature and $k_{\mathrm{B}}$ the Boltzmann constant and where $d=(4\pi{n}/3)^{-1/3}$ is the Wigner-Seitz radius with $n$ the particle density.

The YOCP is a simple versatile model system that behaves as a non-interacting gas, dense liquid or crystalline solid (bcc or fcc) depending on the $(\Gamma,\kappa)$ state point of interest. The YOCP allows the exploration of the full range of potential softness from the long range Coulomb interactions of the one-component plasma (OCP) for $\kappa=0$ to ultra-short range hard-sphere interactions for $\kappa\to\infty$. In spite of concerning a simple purely repulsive divergent-at-the-origin pair interaction potential, its variable softness is one reason that the YOCP is still being actively investigated in statistical mechanics studies. The other main reason is the relevance of the YOCP model to strongly coupled laboratory systems such as complex plasmas\,\cite{morfill2009,bonitz2010,khrapak2012}, charge-stabilized colloidal suspensions\,\cite{boroudjerdi2005}, ultra-cold neutral plasmas\,\cite{killian2007,murillo2015} and even warm dense matter\,\cite{brown1997,wunsch2009}.

On the computer simulation front; detailed thermodynamic, structural and phase equilibria investigations have been carried out with the molecular dynamics (MD)\,\cite{robbins1988,farouki1994,hamaguchi1996,hamaguchi1997}, Monte Carlo (MC)\,\cite{meijer1991,caillol2000,caillol2003} and Langevin dynamics (LD)\,\cite{klumov2010,ott2014,ott2015} techniques. Computational coverage of the entire YOCP phase diagram has resulted in simple accurate parameterizations for the melting line\,\cite{vaulina2000,vaulina2002}, the thermodynamic properties\,\cite{rosenfeld2000a,khrapak2015a,khrapak2015b,khrapak2015c} and the static correlation functions\,\cite{desbiens2016}. On the theoretical front; the thermodynamics of the YOCP have been computed with semi-phenomenological approaches\,\cite{khrapak2014a,khrapak2014b}, whereas thermodynamic and structural properties have been computed with numerous integral equation theory approaches\,\cite{hansen2006}  which include the hypernetted-chain (HNC)\cite{kalman2000}, the soft mean spherical (SMSA)\,\cite{tolias2014,tolias2015}, the variational modified hypernetted-chain (VMHNC)\,\cite{faussurier2004}, the empirically modified hypernetted-chain (EMHNC)\,\cite{daughton2000}, the Rogers-Young (RY)\,\cite{tejero1992} and the isomorph-based empirically modified hypernetted chain approximations (IEMHNC)\,\cite{tolias2019}. Nevertheless, few detailed comparisons between different approximations have been reported thus far.

In this work we present a systematic comparison between different approximate bridge function closures applied to dense YOCP liquids. The objective is to evaluate the performance of the recently-proposed IEMHNC approach that is constructed upon the ansatz of bridge function isomorph invariance\,\cite{tolias2019,lucco2019}. The IEMHNC is a very accurate integral equation theory approach which produces excellent agreement with computer simulations of dense YOCP and bi-Yukawa liquids\,\cite{tolias2019,lucco2019}. The absence of adjustable parameters makes it an ideal candidate for time-consuming computations such as the parametric scans of the phase diagram necessary to obtain equations of state\,\cite{lucco2019}. The comparison is carried out in terms of thermodynamic properties, radial distribution functions, thermodynamic consistency, computational speed and bridge functions
against three advanced established integral equation theory approaches; the VMHNC approach constructed upon the conjecture of bridge function quasi-universality\,\cite{rosenfeld1986}, the mixed RY approach that employs a switching function to interpolate between two fundamental closures\,\cite{rogers1984}, the mixed Ballone-Pastor-Galli-Gazzillo (BPGG) approach that introduces an adjustable parameter to enforce thermodynamic consistency\,\cite{ballone1986}. Possible shortcomings of the IEMHNC bridge function are pinpointed, improvement schemes are proposed and a physically transparent cross-over extension is investigated.

\section{Integral equation theory}\label{sec:theory}

\noindent In case of one-component pair-interacting isotropic systems, the integral equation theory of liquids consists of the Ornstein-Zernike (OZ) equation\,\cite{hansen2006}
\begin{equation}
h(r)=c(r)+n\int c(r')h(|\boldsymbol{r}-\boldsymbol{r}'|)d^3r'\,,\label{eq:theory_oz}
\end{equation}
combined with the following exact non-linear closure condition that is derived from cluster diagram analysis\,\cite{hansen2006}
\begin{equation}
g(r)=\exp\left[-\beta u(r)+h(r)-c(r)+B(r)\right].\label{eq:theory_oz_closure}
\end{equation}
In the above; $g(r)$ denotes the radial distribution function or pair correlation function, $h(r)=g(r)-1$ denotes the total correlation function, $c(r)$ the direct correlation function and $B(r)$ the bridge function. A formally exact expression for the bridge function is required in order to close the system of equations.

Within the diagrammatic framework\,\cite{hansen2006}, bridge diagrams are highly connected suggesting that bridge function evaluations are extremely cumbersome. The bridge function can be formally defined through a virial-type series $B(r)=\sum_{2}^{\infty}d_i(r;T)n^i$, where the unknown functions $d_i(r;T)$ are given by multidimensional integrals involving Mayer functions (f-bond expansion) or through a virial-type series $B(r)=\sum_{2}^{\infty}b_i(r;n,T)n^i$ where the unknown functions $b_i(r;n,T)$ are given by multidimensional integrals involving total correlation functions (h-bond expansion)\,\cite{stell1964,attard1990}. Unfortunately, these expansions converge very slowly already at moderate densities and the coefficient calculations become formidable beyond the third-order term\,\cite{rast1999,kwak2005}. Therefore, approximate closures need to be considered. The bridge function is prescribed either through the radial distribution and direct correlation functions (unique functionality condition) or directly as function of the distance and state variables.

Fundamental closures are formulated on the basis of rigorous approximations of exact infinite functionals and express the bridge function through the indirect correlation function $\gamma(r)=h(r)-c(r)$. The most relevant are the hypernetted-chain (HNC) approach where $B[\gamma(r)]=0$\,\cite{morita1960}, the Percus-Yevick (PY) approach where $B[\gamma(r)]=\ln[1+\gamma(r)]-\gamma(r)$\,\cite{percus1958} and the Martynov-Sarkisov (MS) approach where $B[\gamma(r)]=\sqrt{1+2\gamma(r)}-\gamma(r)-1$\,\cite{martynov1983}. On the other hand, mixed closures either interpolate between or insert adjustable parameters in fundamental closures\,\cite{caccamo1996,bomont2008}. Modified hypernetted-chain closures utilize $B(r)$ parameterizations as function of the distance and the thermodynamic state. Such approximations are usually tailored to specific types of interaction potential, but their applicability can be broadened by introducing correspondence mappings between reference and real systems or between reference states and real states. In the remainder of this section, we shall analyze two modified hypernetted-chain closures and two mixed closures that will later be applied to the YOCP.

\subsection{The isomorph-based empirically modified hypernetted-chain approach}\label{theory_iemhnc}

\noindent The isomorph-based empirically modified hypernetted-chain (IEMHNC) approach is based on the ansatz that bridge functions remain exactly invariant along isomorph curves when expressed in reduced units where the length is normalized to the mean-cubic inter-particle distance $a=n^{-1/3}$ and the energy to $k_{\mathrm{B}}T$\,\cite{tolias2019}. For clarity's sake, we will take a brief detour to introduce isomorph theory.

Isomorphs are phase diagram curves of constant excess entropy, along which a large set of structural as well as dynamic properties are approximately invariant when expressed in the aforementioned reduced units\,\cite{dyre2016}. Isomorphic curves exist only for R-simple systems\,\cite{schroder2014} that, for most practical purposes, can be defined as a broad class of systems whose virial ($W$) and potential energy ($U$) canonical fluctuations are characterized by correlation coefficients $R=\langle\Delta U\Delta W \rangle/\sqrt{\langle(\Delta U)^2\rangle\langle(\Delta W)^2\rangle}\geq0.9$ for an extended portion of the phase diagram\,\cite{gnan2009}. Here $\langle...\rangle$ denotes the NVT ensemble average and $\Delta$ denotes the deviation from the thermodynamic mean. Isomorphs are usually determined via dedicated NVT computer simulations\,\cite{gnan2009}, but their functional form can also be estimated directly from the interaction potential\,\cite{bohling2014}. A recent investigation showed that the YOCP is an R-simple system with exceptionally strong $U-W$ correlations ($R>0.99$) for a large part of the fluid phase\,\cite{veldhorst2015}. In the same work it was concluded that YOCP isomorphs can be approximated to high accuracy by the analytical expression\,\cite{veldhorst2015}
\begin{equation}
\Gamma_{\mathrm{iso}}(\Gamma,\kappa)=\Gamma e^{-\alpha\kappa}\left[1+\alpha\kappa+\frac{(\alpha\kappa)^2}{2}\right]=\mathrm{const},\label{eq:theory_iemhnc_Gamma_iso}
\end{equation}
where $\alpha=a/d=(4\pi/3)^{1/3}$ serves as a conversion factor between the standard plasma literature units where distances are normalized to the Wigner-Seitz radius and the standard liquid theory units where normalization is over the mean-cubic inter-particle distance. This expression is equivalent to the semi-empirical expression\,\cite{vaulina2002} utilized to fit MD data\,\cite{hamaguchi1996,hamaguchi1997} of the YOCP melting line
\begin{equation}
\Gamma_{\mathrm{m}}(\kappa)=\Gamma_{\mathrm{m}}^{\mathrm{OCP}} e^{\alpha\kappa}\left[1+\alpha\kappa+\frac{(\alpha\kappa)^2}{2}\right]^{-1},\label{eq:theory_iemhnc_Gamma_melt}
\end{equation}
where $\Gamma_{\mathrm{m}}^{\mathrm{OCP}}=171.8$ is the OCP coupling parameter at melting\,\cite{hamaguchi1996}. Thus, YOCP isomorphs can be conveniently identified through the ratio $\Gamma/\Gamma_{\mathrm{m}}(\kappa)=\mathrm{const.}$.

We are now in the position to describe the construction of the IEMHNC closure\,\cite{tolias2019}. The approach requires two external inputs: a closed form expression for the isomorphs and a closed-form bridge function expression that is valid along any phase diagram line that possesses a unique intersection point with any isomorph. Such an input allows us to leverage the bridge function invariance ansatz to obtain an expression for the bridge function which is ideally valid for the whole phase diagram, but is practically limited to a phase diagram sub-region due to the restricted validity of the external inputs. The IEMHNC approach can be straightforwardly applied to any system for which such input is available.

For the YOCP, the bridge function has been extracted from MC simulations in the OCP limit ($\kappa=0$) and a simple parametrization has been constructed that is valid for $5.25<\Gamma<171.8$\,\cite{iyetomi1992}. The expression reads as
\begin{align}
B_{\mathrm{OCP}}(r,\Gamma)=&\Gamma\left[-b_0(\Gamma)+c_1(\Gamma)\left(\frac{r}{d}\right)^4+c_2(\Gamma)\left(\frac{r}{d}\right)^6\right.\nonumber\\&\left.+c_3(\Gamma)\left(\frac{r}{d}\right)^8\right]\exp\left[-\frac{b_1(\Gamma)}{b_0(\Gamma)}\left(\frac{r}{d}\right)^2\right].
\label{eq:theory_iemhnc_bOCP}
\end{align}
The functional dependence of the coefficients appearing in $B_{\mathrm{OCP}}(r,\Gamma)$ is provided in Eqs.(21,23) of Ref.\cite{iyetomi1992}. Note that, even though details of the MC extraction have been criticized\,\cite{rosenfeld1992,rosenfeld1996}, the above expression remains the most accurate parameterization available in the literature. In absence of a better alternative, Eq.\eqref{eq:theory_iemhnc_bOCP} can be regarded as an exact representation of the OCP bridge function.

By capitalizing on the OCP bridge function parameterization of Eq.\eqref{eq:theory_iemhnc_bOCP} and the analytical YOCP isomorph expression of Eq.\eqref{eq:theory_iemhnc_Gamma_iso}, the isomorph invariance ansatz can be invoked to obtain the following expression for the YOCP bridge function valid $\forall\kappa$ \& $5.25<\Gamma_{\mathrm{iso}}(\Gamma,\kappa)<171.8$\,\cite{tolias2019}
\begin{equation}
B_{\mathrm{IEMHNC}}(r/d,\Gamma,\kappa)=B_{\mathrm{OCP}}\left[r/d,\Gamma_{\mathrm{iso}}(\Gamma,\kappa)\right].\label{eq:theory_iemhnc_bf}
\end{equation}
The IEMHNC approach has exhibited an excellent agreement with computer simulations in recent applications to YOCP systems\,\cite{tolias2019} and dense bi-Yukawa liquids\,\cite{lucco2019}. In particular, a thorough investigation of the IEMHNC accuracy for the YOCP was conducted where the IEMHNC approach was validated against LD, MD and MC simulation results over the entire dense fluid region\,\cite{tolias2019}. The comparison revealed that the IEMHNC approach possesses remarkable accuracy with predictions of structural properties inside the first coordination cell within $2.0\%$ and of thermodynamic properties within $0.5\%$\,\cite{tolias2019}.

\subsection{The variational modified hypernetted-chain approach}\label{theory_vmhnc}

\noindent The variational modified hypernetted-chain (VMHNC) approach is based on the ansatz of bridge function quasi-universality, which states that bridge functions retain the same functional form irrespective of the interaction potential\,\cite{rosenfeld1979}. This conjecture allows the use of the known bridge function of a well-characterized reference system in the place of the unknown bridge function of the investigated system. The missing ingredient is a mathematical procedure that specifies a one-to-one correspondence between the state points of the two systems. The VMHNC not only uses a convenient reference system but also employs a robust method that leads to an optimal mapping.

In the VMHNC approach, the reference system is the hard-sphere system within the Percus-Yevick approximation (HSPY)\,\cite{rosenfeld1979}. The HS system has a one-dimensional phase diagram where thermodynamic states are entirely specified by the packing fraction $\eta=\pi{n}^3\sigma/6$ with $\sigma$ the hard-sphere diameter, whereas the HSPY system possesses an exact analytical bridge function that stems from the Wertheim-Thiele solution\,\cite{wertheim1963,thiele1963}. It can be proven that the optimal effective packing fraction $\eta_{\mathrm{eff}}$ ensuring that $B_{\mathrm{HSPY}}(r;\eta_{\mathrm{eff}})\approx{B}(r;n,T)$ is acquired by minimizing the VMHNC free energy functional\,\cite{rosenfeld1986}. Hence, the VMHNC closure becomes\,\cite{rosenfeld1979,rosenfeld1986}
\begin{equation}
g(r)=\exp\left[-\beta{u}(r)+h(r)-c(r)+B_{\mathrm{HSPY}}(r;\eta_{\mathrm{eff}})\right]\label{eq:theory_vmhnc_bf}
\end{equation}
and the effective packing fraction $\eta_{\mathrm{eff}}$ is recovered from
\begin{equation}
\frac{d{\delta}_\phi}{d\eta}-\frac{n}{2}\int\left[g(r)-g_{\mathrm{HSPY}}(r;\eta)\right]\frac{\partial{B}_{\mathrm{HSPY}}(r;\eta)}{\partial\eta}d^3r=0\,,\label{eq:theory_vmhnc_minimization}
\end{equation}
with $\delta_\phi$ the difference between the Carnahan-Starling\,\cite{carnahan1969} and PY-virial free energies, which reads as
\begin{equation}
\delta_\phi(\eta)=\frac{\eta(4-3\eta)}{(1-\eta)^2}-\frac{6\eta}{(1-\eta)}-2\ln(1-\eta),
\end{equation}
with ${B}_{\mathrm{HSPY}}(r;\eta)$ the HSPY bridge function given by\,\cite{rosenfeld1979}
\begin{equation}
B_{\mathrm{HSPY}}(r;\eta)=
\begin{cases}
\ln[-c_{\mathrm{HSPY}}(r;\eta)]+c_{\mathrm{HSPY}}(r;\eta)+1,r\leq\sigma \\
\ln[g_{\mathrm{HSPY}}(r;\eta)]-g_{\mathrm{HSPY}}(r;\eta)+1,\,\,\,\,r>\sigma
\end{cases}
\label{eq:theory_vmhnc_bfhspy}
\end{equation}
where both the direct correlation function $c_{\mathrm{HSPY}}(r)$ and the radial distribution function $g_{\mathrm{HSPY}}(r)$ are analytically accessible. We should point out that $c_{\mathrm{HSPY}}(r)$ follows a closed-form expression valid at all distances, but only the $g_{\mathrm{HSPY}}(r)$ Laplace transform is known analytically\,\cite{wertheim1963}. The complexity of the Laplace inversion necessary to obtain $g_{\mathrm{HSPY}}(r)$ becomes unwieldy at large distances and an exact analytical expression for the radial distribution function is only available for $r<6\sigma$\,\cite{smith1970,henderson1978}. This issue does not constitute a limitation, since accurate extrapolation techniques based on exact asymptotic limits can be employed to extend $g_{\mathrm{HSPY}}(r)$ beyond $r=6\sigma$, see sec.\ref{sec:numerics}.

It should be noted that the VMHNC virial-energy consistency is automatically satisfied without any artificial enforcement and that the VMHNC virial-compressibility consistency is optimized via the state correspondence rule\,\cite{rosenfeld1986}. The VMHNC approximation has shown excellent agreement with computer simulations of diverse pair-interacting systems including inverse power law (IPL) liquids\,\cite{laird1990}, YOCP liquids\,\cite{laird1990,faussurier2004}, hard core Yukawa fluids\,\cite{caccamo1999}, Lennard-Jones binary mixtures\,\cite{gonzalez1992}, binary ionic mixtures\,\cite{kahl1996}, HS mixtures\,\cite{kahl1996} and penetrable sphere systems\,\cite{rosenfeld2000}. The VMHNC approach has also been successfully complemented with effective potentials and employed to predict the thermodynamic \& static properties of liquid metals\,\cite{chen1992,gonzalez1993,bhuiyan1996} and warm dense matter\,\cite{starrett2015,daligault2016}.

For the sake of completeness, we mention the reference hypernetted-chain (RHNC) approach that is also based on bridge function quasi-universality but selects the exact HS system as reference\,\cite{lado1982,lado1983}. Since now ${d{\delta}_\phi}/{d\eta}=0$, the effective packing fraction $\eta_{\mathrm{eff}}$ equation becomes
\begin{equation}
\frac{n}{2}\int\left[g(r)-g_{\mathrm{HS}}(r;\eta)\right]\frac{\partial B_{\mathrm{HS}}(r;\eta)}{\partial\eta}d^3r=0.
\end{equation}
Here; the radial distribution function $g_{\mathrm{HS}}(r)$ is given by the Verlet-Weis-Henderson-Grundke parametrization\,\cite{verlet1972,henderson1975}, whereas the bridge function $B_{\mathrm{HS}}(r)$ can either be obtained from $B_{\mathrm{HS}}(r)=\ln[y_{\mathrm{HS}}(r)]-\gamma_{\mathrm{HS}}(r)$ with $\ln[y_{\mathrm{HS}}(r)]$ the cavity function that has been parameterized as a cubic polynomial\,\cite{henderson1975} and $\gamma_{\mathrm{HS}}(r)$ the indirect correlation function acquired via numerical Fourier transforms\,\cite{henderson1975}, or can be expressed with the parametrization available in Refs.\cite{malijevsky1987,labik1989}. Naturally, there are strong similarities between the VMHNC and RHNC approaches that are expected to possess the same level of accuracy in thermodynamic properties\,\cite{rosenfeld1986}. However, the RHNC is known to produce a small un-physical dip around the second maximum of the radial distribution function that is caused by the $g_{\mathrm{HS}}(r)$ input\,\cite{lado1983}. Although this dip does not seem to have a detrimental effect on thermodynamic properties, it can be entirely avoided within the VMHNC approach.

With respect to the present study, it is worth bringing forth the detailed YOCP investigation with the VMHNC approach carried out by Faussurier\,\cite{faussurier2004}. The author performed an extensive study of phase equilibria, thermodynamic, static and dynamic properties, but did not report any comparison with other integral equation theories.

\subsection{The Rogers-Young approach}\label{sec:theory_ry}

\noindent In case of purely repulsive potentials, the PY and HNC radial distribution functions have been demonstrated to bracket the exact radial distribution functions. For thermodynamically stable systems, it is also known that the bridge functions are particularly short-ranged rapidly decaying to zero. These observations led to the formulation of the Rogers-Young (RY) approach, a mixed approximation that interpolates continuously between the PY closure at short distances and the HNC closure at large distances\,\cite{rogers1984}. The interpolation is realized with the aid of a switching function that contains an adjustable parameter which is determined by imposing thermodynamic consistency between the virial and compressibility routes to the equation of state. The bridge function reads as\,\cite{rogers1984}
\begin{equation}
B_{\mathrm{RY}}[\gamma(r)]=\ln\left[1+\frac{\exp[\gamma(r)f(r)]-1}{f(r)}\right]-\gamma(r)\,,\label{eq:theory_ry_bf}
\end{equation}
with $f(r)=1-\exp{\left(-s{r}\right)}$ the switching function and $s$ the adjustable parameter. Other expressions can be utilized for the monotonically increasing switching function, provided that it is bounded between zero and unity.

The RY approximation is particularly suitable for HS and IPL ($n\geq4$) systems, for which it has been observed that the adjustable parameter is nearly thermodynamic state independent which implies a drastic reduction in computational cost\,\cite{rogers1984}. However, this holds neither for the OCP\,\cite{rogers1984} nor for the YOCP\,\cite{tejero1992} where the optimization procedure needs to be repeated at each state point. The RY approach has been successfully applied to various, predominantly repulsive, pair interactions including ultra-soft logarithmic core potentials\,\cite{watzlawek1998}, square shoulder potentials\,\cite{lang1999}, the gaussian core model\,\cite{lang2000}, isotropic core-softened potentials\,\cite{barros2006}, the repulsive Jagla potential\,\cite{kumar2005} as well as hard core Yukawa fluids\,\cite{heinen2011}. Note that the RY optimization routine does not converge for potentials possessing an attractive tail. In such cases, the PY branch should be substituted with a SMSA branch\,\cite{zerah1986}.

We point out that the RY approximation has been previously applied to YOCP systems. Tejero \emph{et al.} demonstrated that the RY excess pressures exhibit an excellent agreement with MC simulations for intermediate and high screening parameters ($\kappa\geq1.8$)\,\cite{tejero1992}, whereas Gapinski \emph{et al.} studied the RY freezing lines with the one-phase Hansen-Verlet freezing rule focusing on charge-stabilized colloidal applications\,\cite{gapinski2012}. The present investigation will attempt a more extended comparison with computer simulations at the level of radial distribution functions.

\subsection{The Ballone-Pastore-Galli-Gazzillo approach}\label{sec:theory_bpgg}

\noindent For HS systems, the MS closure performs considerably better than the PY closure\,\cite{martynov1983}, albeit it cannot be solved analytically. However, it is afflicted by the non-existence of physical solutions in case of dilute systems, where the square root argument might attain negative values leading to a complex bridge function. The Ballone-Pastore-Galli-Gazzillo (BPGG) approach is a mixed approximation that generalizes the MS closure by introducing a free parameter adjusted by imposing thermodynamic consistency between the virial and compressibility routes to the equation of state\,\cite{ballone1986}. The bridge function reads as\,\cite{ballone1986}
\begin{equation}
B_{\mathrm{BPGG}}[\gamma(r)]=\left[1+s\gamma(r)\right]^{1/s}-\gamma(r)-1\,,\label{eq:theory_bpgg_bf}
\end{equation}
where $1\leq{s}\leq2$ is the adjustable parameter. The BPGG approach clearly mixes the MS closure ($s=2$) with the HNC closure ($s=1$). The $s$ value can be approximated with a rational number of odd nominator with arbitrary precision, eliminating the possibility of complex values.

The BPGG approach has been mainly applied to one-component HS systems as well as to binary and ternary HS mixtures with additive or non-additive diameters\,\cite{ballone1986,kjellander1989,biben1991,jung1995,gazzillo1995}. To our knowledge, the BPGG approach has never been applied to the YOCP.

\section{Numerical implementation}\label{sec:numerics}

We have numerically solved Eqs.(\ref{eq:theory_oz},\ref{eq:theory_oz_closure}) together with each of the four bridge function closures discussed in sec. \ref{sec:theory}. Given that the solution of each integral theory approximation possesses its own peculiarities, we shall address them separately in what follows starting from the least numerically complicated.

\subsection{The IEMHNC algorithm}\label{sec:numerics_emhnc}

\noindent The IEMHNC algorithm has the simplest structure, since the approach involves neither correspondence mapping nor consistency enforcement. The algorithm solves Eqs.(\ref{eq:theory_oz},\ref{eq:theory_oz_closure}) with the bridge function of Eq.\eqref{eq:theory_iemhnc_bf} for any $(\Gamma,\kappa)$ combination. Essentially, the IEMHNC algorithm is also embedded in the VMHNC, RY and BPGG algorithms. The equations are solved with Picard iterations in Fourier space. Mixing techniques are utilized to facilitate convergence\,\cite{duh1995l} and well-established long-range decomposition techniques\,\cite{ng1974,faussurier2004} are employed within the weak screening regime ($\kappa<1.0$). Fourier transforms are computed with efficient FFT algorithms over a domain extending up to $R_{\mathrm{max}}=20d$ with real space resolution $\Delta r=10^{-3}d$ and reciprocal space resolution $\Delta k =\pi /R_{\mathrm{max}}$. The convergence criterion for the Picard iterations is formulated in terms of the indirect correlation function $\gamma$ in Fourier space and reads as $|\gamma_m(k)-\gamma_{m-1}(k)|<10^{-5}$ $\forall k$.

\subsection{The VMHNC algorithm}\label{sec:numerics_vmhnc}

\begin{figure}
	\centering
	\includegraphics[width=3.4in]{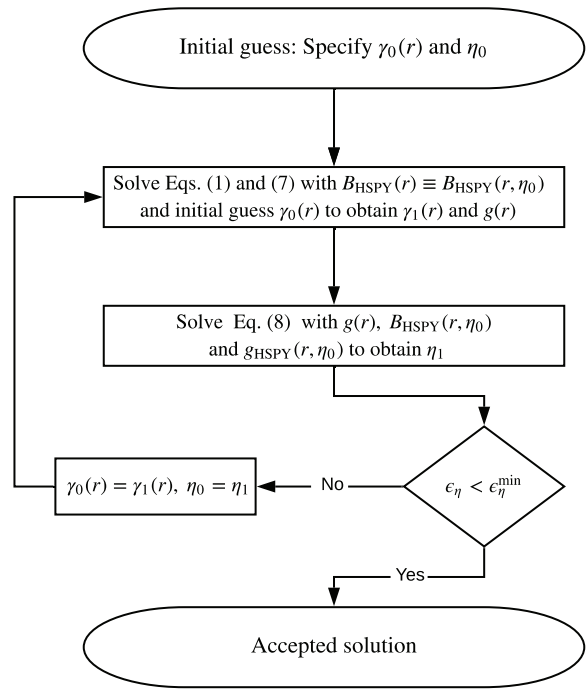}
	\caption{Structure of the algorithm employed for the numerical solution of the VMHNC approximation. Here $\gamma(r)$ denotes the indirect correlation function, $\eta$ the packing fraction and the relative error $\epsilon_{\eta}$ used to monitor the packing fraction convergence is defined as $\epsilon_\eta=|\eta_0-\eta_1|/\eta_1$.}\label{fig:numerics_vmhnc_algorithm}
\end{figure}

\noindent Compared to the IEMHNC, the VMHNC algorithm contains an additional iteration cycle dedicated to the identification of the optimal effective packing fraction. The structure of the VMHNC algorithm is summarized in Fig.\ref{fig:numerics_vmhnc_algorithm}.  A double initial guess is required for the structural properties and the packing fraction $\eta_0$. Eqs.(\ref{eq:theory_oz},\ref{eq:theory_vmhnc_bf}) are then solved with the Picard iteration method described above. This leads to an estimate for the radial distribution function $g(r)$ that is employed in Eq.\eqref{eq:theory_vmhnc_minimization} to acquire an updated packing fraction value. This updated $\eta_1$ value is then used to compute the relative error $\epsilon_\eta=|\eta_0-\eta_1|/\eta_1$ and the procedure is repeated until the convergence criterion $\epsilon_\eta<10^{-5}$ is satisfied, at each cycle updating $\eta_0$ and the structural properties with the most recent estimates.

The optimization equation, Eq.\eqref{eq:theory_vmhnc_minimization}, is solved with standard root-finding algorithms that are based on Newton's method. The required first order derivative of the bridge function is approximated with the central finite difference with a $\Delta\eta=10^{-3}\eta$ resolution.
The analytical solution of $g_{\mathrm{HSPY}}(r)$ is extrapolated for $r\geq6\sigma$ with the long-range fit $g_{\mathrm{fit}}(r)=[A_0(\eta)/r]\exp\left[-A_1(\eta)r\right]\sin[A_2(\eta)r+A_3(\eta)]+1$\,\cite{kolafa2002}. Analytical expressions for the set of coefficients $\{A_0,A_1,A_2,A_3\}$ valid for $0.1\leq6\eta/\pi\leq1.0$ are given in Eqs.(14-17) of Ref.\cite{kelly2016}. This analytical fit is the theoretical asymptotic limit of the radial distribution function above the Fisher-Widom line\,\cite{evans1994,carvalho1999}. The accuracy of the fit in the extrapolated range has been verified after comparisons with numerical Laplace inversions with the robust Durbin method\,\cite{tolias2014}.

The present implementation of the VMHNC approach was successfully validated against the static structure factor first peak results reported in Table I of Ref.\cite{faussurier2004}. It is worth noting that more stringent convergence criteria for the packing fraction ($\epsilon_\eta<10^{-7}$) and higher resolution in the finite difference estimate of the bridge function derivative ($\Delta\eta=10^{-5}\eta$) barely affect the final results.

\subsection{The RY and BPGG algorithms}\label{sec:numerics_ry+bpgg}

\noindent The algorithm for the solution of the RY \& BPGG approaches bears a similar structure to the VMHNC algorithm illustrated in Fig.\ref{fig:numerics_vmhnc_algorithm}. The main difference lies on the replacement of the variational equation with the thermodynamic consistency requirement that is formulated in terms of reduced excess inverse isothermal compressibilities with added neutralizing background contributions, $\mu^{\mathrm{p}}$. From the statistical and virial routes, we have\,\cite{tolias2019}
\begin{align}
&\mu_{\mathrm{stat}}^{\mathrm{p}}(\Gamma,\kappa;s)=-n\int\left[c(r)+\beta u(r)\right]d^3r\,,\nonumber\\
&\mu_{\mathrm{vir}}^{\mathrm{p}}(\Gamma,\kappa;s)=p_{\mathrm{ex}}^{\mathrm{p}}(\Gamma,\kappa)+\frac{\Gamma}{3}\frac{\partial{p}_{\mathrm{ex}}^{\mathrm{p}}(\Gamma,\kappa)}{\partial\Gamma}-\frac{\kappa}{3}\frac{\partial{p}_{\mathrm{ex}}^{\mathrm{p}}(\Gamma,\kappa)}{\partial\kappa}\,,\nonumber
\end{align}
where $p_{\mathrm{ex}}^{\mathrm{p}}=-(1/6)\beta n\int r(du/dr)[g(r)-1]d^3r$ is the reduced excess pressure (particle-particle plus particle-background interactions). The $s$-dependence stems from the implicit $s-$dependence of the bridge function in the RY and BPGG approaches, implying the same for $g(r)$ and $c(r)$. The thermodynamic consistency requirement simply reads as $\mu_{\mathrm{stat}}^{\mathrm{p}}(\Gamma,\kappa;s)=\mu_{\mathrm{vir}}^{\mathrm{p}}(\Gamma,\kappa;s)$.

An initial guess for the static correlations and the free parameter $s_0$ is utilized to solve Eqs.(\ref{eq:theory_oz},\ref{eq:theory_oz_closure}) with the bridge function of Eq.\eqref{eq:theory_ry_bf} or Eq.\eqref{eq:theory_bpgg_bf}. Afterwards, an updated version of the free parameter, $s_1$, is obtained by imposing thermodynamic consistency and used to compute the error $\epsilon_s=|s_0-s_1|/s_1$. The solution is accepted when $\epsilon_s<10^{-5}$, otherwise the computational cycle is repeated. A simple secant method has been employed for the solution of the consistency requirement. It should be noted that the determination of the virial compressibility is a particularly time-consuming operation, since the second order excess pressure derivatives are computed with central finite differences for $\Delta\Gamma=10^{-3}\Gamma$ and $\Delta\kappa=10^{-3}\kappa$, which requires the numerical solution of the integral equation system at five state points for any $(\Gamma,\kappa)$ pair. The present implementation of the RY approach was successfully validated against YOCP excess pressure results reported in Table I of Ref.\cite{tejero1992}.

\section{Numerical results}\label{sec:results}

\noindent The IEMHNC, VMHNC, RY and BPGG approximations have been compared for an extensive set of YOCP state points covering the dense fluid region $0.1\leq\Gamma/\Gamma_{\mathrm{m}}(\kappa)<1.0$ and ranging from the infinitely long-ranged OCP with $\kappa=0$ to the short range interactions arising for $\kappa=5$. The performances of these four integral equation theory approaches have been compared in terms of radial distribution functions, thermodynamic properties, thermodynamic consistency and computational cost.

\subsection{Radial distribution functions} \label{sec:results_rdf}

\begin{figure}[htbp]
\centering
\includegraphics[width=2.94in]{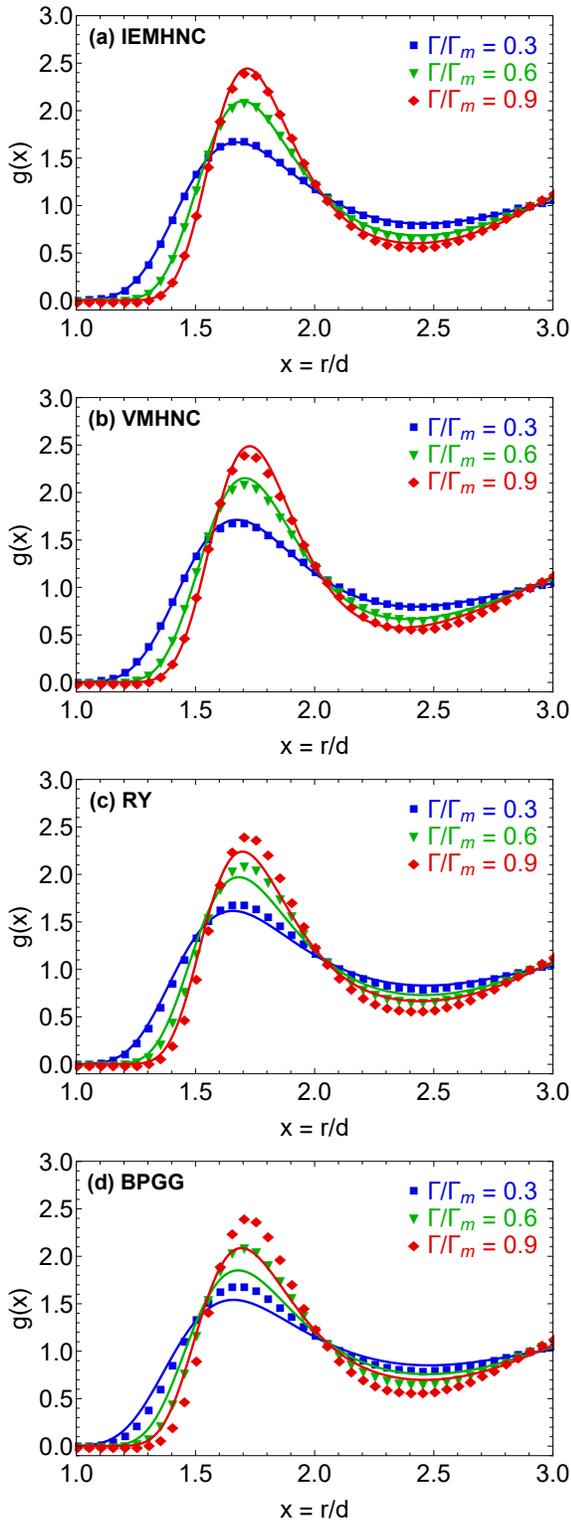}
\caption{Radial distribution functions resulting from MD simulations (discrete points) and different integral equation theory approaches (solid lines), namely; (a) the IEMHNC, (b) the VMHNC (c) the RY, (d) the BPGG. Results for $\kappa=1.0$ and $\Gamma/\Gamma_{\mathrm{m}}=0.3$ (blue), 0.6 (green), 0.9 (red). The melting point coupling parameter $\Gamma_{\mathrm{m}}$ was computed from the semi-empirical expression of Eq.\eqref{eq:theory_iemhnc_Gamma_melt} resulting in $\Gamma_{\mathrm{m}}=220.18$. The MD results were down-sampled in order to improve visibility.}\label{fig:results_rdf_kappa1.0}
\end{figure}

\begin{figure}[htbp]
\centering
\includegraphics[width=2.94in]{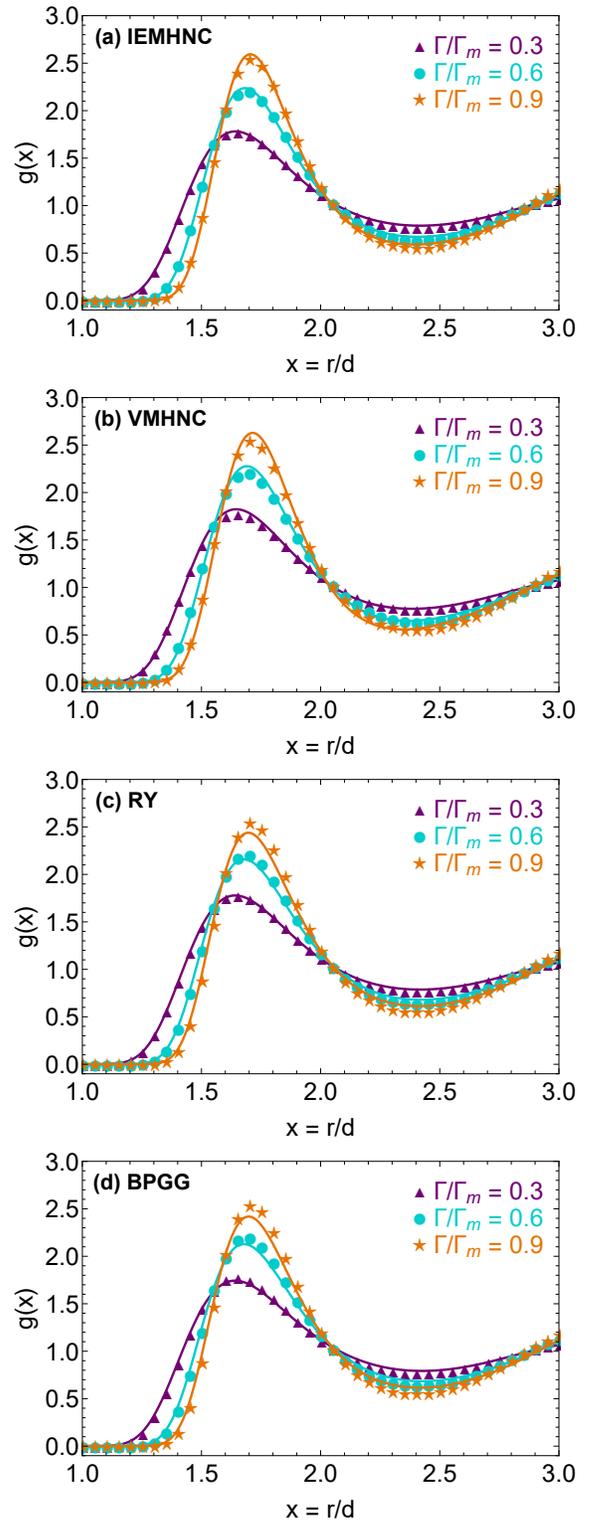}
\caption{Radial distribution functions resulting from MD simulations (discrete points) and different integral equation theory approaches (solid lines), namely; (a) the IEMHNC, (b) the VMHNC (c) the RY, (d) the BPGG. Results for $\kappa=3.0$ and $\Gamma/\Gamma_{\mathrm{m}}=0.3$ (purple), 0.6 (cyan), 0.9 (orange). The melting point coupling parameter $\Gamma_{\mathrm{m}}$ was computed from the semi-empirical expression of Eq.\eqref{eq:theory_iemhnc_Gamma_melt} resulting in $\Gamma_{\mathrm{m}}=1234.51$. The MD results were down-sampled in order to improve visibility.}\label{fig:results_rdf_kappa3.0}
\end{figure}

\noindent A graphical comparison between the radial distribution functions resulting from the integral equation theory approaches with the \enquote{exact} radial distribution functions extracted from MD simulations is illustrated in Figs.\ref{fig:results_rdf_kappa1.0},\ref{fig:results_rdf_kappa3.0} for different screening lengths and coupling parameters. The MD simulations were performed with the LAMMPS package\,\cite{plimpton1995} using 8000 particles in the canonical ensemble and truncating the interaction potential at a distance of $10d$. It is apparent that the IEMHNC and the VMHNC approximations, see panels (a) and (b), accurately reproduce the MD results. On the other hand, the mixed RY and BPGG approaches, see the panels (c) and (d), clearly exhibit appreciable deviations around the first maximum and the first non-zero minimum. Furthermore, it can also be observed that the IEMHNC and VMHNC approaches seem to retain the same level of accuracy regardless of the value of the screening parameter, while the self-consistent methods become noticeably less accurate closer to the OCP limit. In this regard, compare the panels (c) and (d) of Fig.\ref{fig:results_rdf_kappa1.0} with their Fig.\ref{fig:results_rdf_kappa3.0} counterpart. Therefore, already from this first qualitative comparison we can conclude that the VMHNC and IEMHNC approaches produce far more accurate results than those obtained with the mixed approaches that enforce thermodynamic consistency.

For more quantitative comparison of the integral equation theory approaches, we resort to some key functional properties of the radial distribution function that have been recently extracted in a comprehensive LD investigation of the weakly screened YOCP\,\cite{ott2015}, which probed three values of the screening parameter ($\kappa=0,1,2$) in combination with coupling parameters between unity and crystallization. The functional characteristics extracted from LD simulations include the edge of the correlation void (estimated as the position where $g[r]=0.5$), the magnitude as well as the position of
the first maximum, the first non-zero minimum and the second maximum. Extensive tabulations of the radial distribution function characteristics as obtained from LD simulations and from the IEMHNC, VMHNC, RY, BPGG and HNC approximations (the latter added as reference) are provided in the supplementary material\,\cite{supplementary}. In what follows, we will mainly focus on the first coordination cell.

The overall conclusion for the IEMHNC and VMHNC approaches is that they are almost indistinguishable from simulations and produce similar estimates for the YOCP structural properties. Both are capable of predicting the edge of the correlation void and the first maximum position with deviations always less than $1.5\%$. Concerning the magnitude of the first maximum, the VMHNC approach produces superior estimates for $\Gamma/\Gamma_{\mathrm{m}}\leq0.4$, but it can't match the 1.0\% accuracy that is achieved by the IEMHNC approach at higher coupling parameters. The most noticeable deviations between the IEMHNC and the VMHNC approximations are observed around the first non-zero minimum, where the VMHNC approach is accurate within $3\%$ both in magnitude and position while the IEMHNC approach, albeit maintaining a comparable accuracy for the position, exhibits deviations as large as $6\%$ for the magnitude. These observations hold for any value of $\kappa\leq2.0$, thereby confirming the observation that the accuracy of the IEMHNC and VMHNC approaches is independent of the screening parameter.

Proceeding with the less accurate RY and BPGG approaches, it can be easily concluded that they are outperformed by the IEMHNC and VMHNC approaches. This becomes particularly clear when inspecting the magnitude of the first maximum for state points that lie close to crystallization. For such state points, the RY and BPGG approaches deviate more than $15\%$ from the LD results. Finally, as already observed in Figs.\ref{fig:results_rdf_kappa1.0},\ref{fig:results_rdf_kappa3.0}, the performance of both mixed integral equation theory approaches worsens when approaching the OCP limit with the BPGG approach practically reducing to the HNC approach.

\subsection{Thermodynamic properties} \label{sec:results_thermo_properties}

\noindent Comprehensive MD results for the reduced excess internal energy $u_{\mathrm{ex}}(\Gamma,\kappa)=\frac{1}{2}\beta n \int u(r)g(r)d^3r$ are available in the literature\,\cite{farouki1994,hamaguchi1996,hamaguchi1997}. It should be noted that the above internal energy expression includes only the contribution from particle-particle interactions and leaves out the contribution of the neutralizing background. Tabulations of $u_{\mathrm{ex}}$ obtained from MD simulations and from the IEMHNC, VMHNC, RY, BPGG and HNC approximations (the latter added as a reference) are available in the supplementary material\,\cite{supplementary}.

The main conclusions obtained from the comparison in the dense fluid region are the following. For small to intermediate screening parameters $(\kappa<3.0)$, the VMHNC approach produces the best estimates for the internal energies with deviations from the computer simulation results that are always smaller than $0.1\%$. The IEMHNC approach, albeit not as accurate as the VMHNC, exhibits minor deviations that do not exceed $0.5\%$ and is far more accurate than the RY and BBPG approaches. For large screening parameters $(\kappa\geq3.0)$, all approximations become comparable: the IEMHNC approximation retains an accuracy of 0.5\% , while the VMHNC, RY and BPGG approaches are accurate within 1.5\% with the highest deviations observed close to the melting point.

The RY and BPGG approaches are characterized by acceptable predictions for the YOCP internal energies even though they exhibit large deviations in the YOCP radial distribution functions. This is a clear manifestation of the smoothing role that integral operators, used to compute the thermodynamic properties from the radial distribution function, can play on $g(r)$ imperfections.

\subsection{Thermodynamic consistency} \label{sec:results_thermo_consistency}

\noindent The IEMHNC and VMHNC approaches have been further analyzed to assess to which extent statistical-virial thermodynamic consistency $\mu_{\mathrm{stat}}^{\mathrm{p}}(\Gamma,\kappa)=\mu_{\mathrm{vir}}^{\mathrm{p}}(\Gamma,\kappa)$ is satisfied. The RY and BPGG approaches were naturally excluded from this investigation, since they both numerically enforce thermodynamic consistency.

Extensive tabulations of $\mu_{\mathrm{stat}}^{\mathrm{p}}$ and $\mu_{\mathrm{vir}}^{\mathrm{p}}$ resulting from the IEMHNC and the VMHNC approximations are also reported in the supplementary material\,\cite{supplementary}. The investigated state points feature five values of the screening parameters $\kappa=\{1,2,3,4,5\}$ and a wide range of coupling parameters. The overall conclusion is that both the IEMHNC and VMHNC approaches exhibit a high degree of thermodynamic consistency with deviations between the statistical and the virial paths to the compressibility which are always smaller than $15\%$ for the IEMHNC approach and smaller than $8\%$ for the VMHNC approach. Both approaches show the highest deviations in the OCP limit where the VMHNC deviations peak at low coupling $(\Gamma/\Gamma_{\mathrm{m}}=0.2)$, while the largest IEMHNC deviations are observed close to melting. When $\kappa\geq4.0$, both approximations lead to comparable deviations which are smaller than 1.0\% for any value of the coupling parameter.

\begin{table}
	\caption{The computational cost of different integral equation theory approaches when applied to dense YOCP liquids. The selected state points are provided in the first two columns. Columns four to six report the CPU time required for the convergence of the VMHNC, RY, BPGG algorithms normalized to the CPU time consumed to solve the IEMHNC approach. In all four numerical routines, the same structural initial guess (the HNC solution) has been employed.}\label{tab:results_time}
	\centering
	\begin{tabular}{cccccc}
		\hline
		$\kappa$ & $\Gamma$ & $\Gamma/\Gamma_{\mathrm{m}}$ &  VMHNC             & RY                & BPGG              \\ \hline
		0.0      &    10    &       0.06                   &  37.9              & 42.2              & 18.4              \\
		0.0      &    40    &       0.23                   &  30.7              & 26.6              & 12.1              \\
		0.0      &    80    &       0.47                   &  23.1              & 18.2              & 5.99              \\
		0.0      &   120    &       0.70                   &  18.8              & 16.0              & 7.11              \\
		0.0      &   160    &       0.93                   &  15.9              & 13.2              & 5.79              \\
		1.0      &    15    &       0.07                   &  68.7              & 28.2              & 16.6              \\
		1.0      &    50    &       0.23                   &  36.7              & 20.8              & 9.39              \\
		1.0      &   105    &       0.48                   &  23.3              & 15.9              & 7.71              \\
		1.0      &   155    &       0.71                   &  19.0              & 14.6              & 9.23              \\
		1.0      &   200    &       0.93                   &  16.6              & 13.5              & 7.32              \\
		2.0      &    30    &       0.07                   &  84.3              & 37.5              & 21.5              \\
		2.0      &   100    &       0.23                   &  33.2              & 22.5              & 13.9              \\
		2.0      &   210    &       0.48                   &  22.9              & 19.8              & 13.0              \\
		2.0      &   310    &       0.70                   &  19.1              & 18.2              & 10.8              \\
		2.0      &   400    &       0.91                   &  16.8              & 17.0              & 7.02              \\ \hline
	\end{tabular}
\end{table}

\subsection{Computational cost} \label{sec:results_time}

\noindent Apart from the theoretical insight they provide, integral equation theory methods also retain a practical advantage over computational equilibrium methods, since they enable access to the liquid structural and thermodynamic properties in a small fraction of the time that is required for computer simulations. Therefore, in detailed comparisons of different integral equation theory approaches, a thorough assessment of both accuracy and computational speed should determine which method is most suitable to be utilized in place of simulations.

Amongst the four integral equation theory approaches investigated in this work, it should be expected that the IEMHNC approach is by far the least expensive in terms of computational resources, since it involves neither an optimization procedure nor thermodynamic consistency enforcement. Aiming to quantify the computational cost, we have monitored the time required to numerically solve the IEMHNC, VMHNC, RY and BPGG approaches.  The normalized CPU time has been reported in Table \ref{tab:results_time} for a selection of YOCP state points. For an objective comparison of the CPU consumption, all approximations have been solved starting from the same initial guess for the structural properties. In particular, for each $(\Gamma,\kappa)$ state point that is reported in Table \ref{tab:results_time}, the HNC approach was first solved and the HNC solution was then employed as the initial guess for each of the four approaches.

As expected, the IEMHNC approach is revealed to be much less time demanding. To be concrete, the IEMHNC approximation converges $15-84$ times faster than the VMHNC approach, $13-42$ times faster than the RY approach and $5-21$ times faster than the BPGG approach. It is worth noting that, regardless of the screening parameter, the relative computational cost of the VMHNC approach decreases as the coupling parameter increases. The same applies for the RY approach, albeit in a quite less pronounced manner.

\section{Discussion}

\subsection{On the conjecture of the isomorph invariance of bridge functions}\label{sec:discussion_vmhnc_invariance}

\noindent The ansatz of the exact isomorph invariance of bridge functions has been formulated upon a number of general observations\,\cite{tolias2019}. The high accuracy of the IEMHNC results obtained in the entire YOCP dense liquid phase together with the fact that the IEMHNC approach retains the same level of accuracy regardless of the screening parameter, indirectly support the bridge function isomorph-invariance conjecture upon which the IEMHNC has been built\,\cite{tolias2019,lucco2019}.  As we shall demonstrate in what follows, additional indirect support to the invariance conjecture stems from the new observation that the VMHNC bridge function for the YOCP is also isomorph invariant.

HS thermodynamic, structural and dynamical properties are entirely specified by the packing fraction $\eta$. This applies also to the VMHNC bridge function, $B_{\mathrm{HSPY}}(r;\eta)$. Therefore, in order to the show that the VMHNC bridge function is isomorph invariant for YOCP systems, it only suffices to demonstrate that the effective packing fraction which maps the YOCP to the HSPY remains constant along any YOCP isomorph curve.

\begin{figure}[htbp]
	\centering
	\includegraphics[width=3.2in]{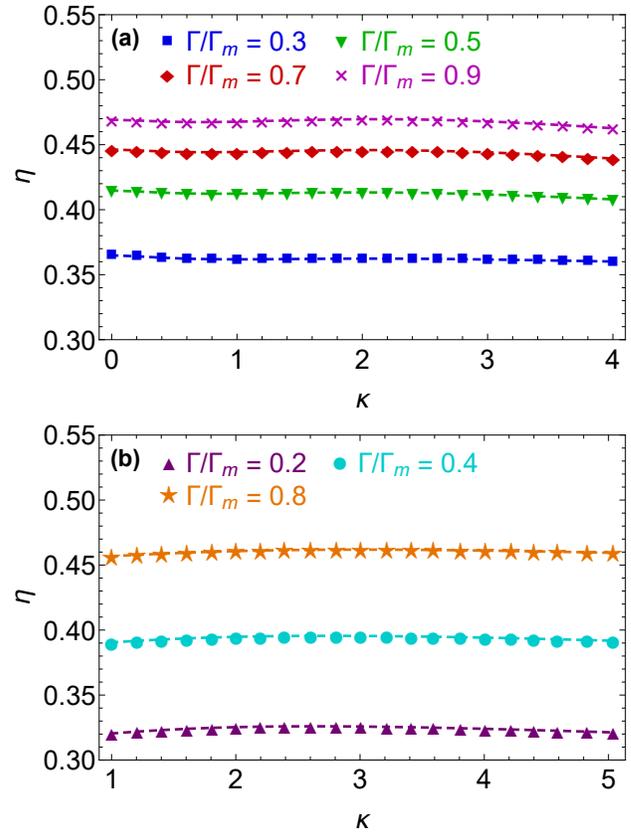}
	\caption{Effective packing fraction obtained by solving the VMHNC approach for isomorphic YOCP state points. Panel (a): Twenty-one screening parameter values within $\kappa\in[0,4]$ with the corresponding coupling parameters $\Gamma$ obtained from the analytical approximation $\Gamma/\Gamma_{\mathrm{m}}=\mathrm{const}$. Four isomorph curves are considered, namely; $\Gamma/\Gamma_{\mathrm{m}}=0.3$ (squares), 0.5 (triangles), 0.7 (diamonds), 0.9 (crosses). Panel (b): Twenty-one screening parameter values within $\kappa\in[1,5]$ with the corresponding coupling parameters $\Gamma$ obtained via the direct isomorph check. Three isomorph curves are considered that can be well approximated by $\Gamma/\Gamma_{\mathrm{m}}=0.2$ (triangles), 0.4 (circles), 0.8 (stars). Dashed lines have been added to guide the eye.}
	\label{fig:discussion_vmhnc_invariance}
\end{figure}

We have considered seven isomorph curves covering the dense fluid region from $\Gamma/\Gamma_{\mathrm{m}}=0.2$ up to $\Gamma/\Gamma_{\mathrm{m}}=0.9$. We have employed two methods in order to compute the state point coordinates belonging to the same YOCP isomorph starting from a reference state point at $(\Gamma_0,\kappa_0)$. The approximate method utilizes the analytical expression $\Gamma/\Gamma_{\mathrm{m}}(\kappa)=\Gamma_0/\Gamma_{\mathrm{m}}(\kappa_0)$ and is employed to trace four isomorphs $\Gamma/\Gamma_{\mathrm{m}}=\{0.3,0.5,0.7,0.9\}$ within $\kappa\in[0,4]$. The exact method requires special MD simulations with the direct isomorph check\,\cite{gnan2009} and is employed to trace three isomorphs $\Gamma/\Gamma_{\mathrm{m}}\approx\{0.2,0.4,0.8\}$ within $\kappa\in[1,5]$. The present NVT MD simulations featured $8192$ particles with the interaction potential truncated at $10a$. We avoided the direct isomorph check near the OCP-limit, since the aforementioned approximate method should be very accurate in the range $\kappa\in[0,1]$. The effective YOCP packing fractions for those seven sets of isomorphic state points are illustrated in Fig.\ref{fig:discussion_vmhnc_invariance}. It is evident that $\eta$ barely changes along any isomorph confirming that the VMHNC bridge function for the YOCP is isomorph invariant. It is also worth noting the entire YOCP dense liquid region is covered by effective packing fractions between $\eta=0.30$ and $\eta=0.48$. The latter observation can be exploited to improve initial guesses in the VMHNC numerical routine.

\subsection{Integral equation theory comparison at the level of bridge functions}\label{sec:discussion_bridge function}

\begin{figure}
\centering
\includegraphics[width=3.0in]{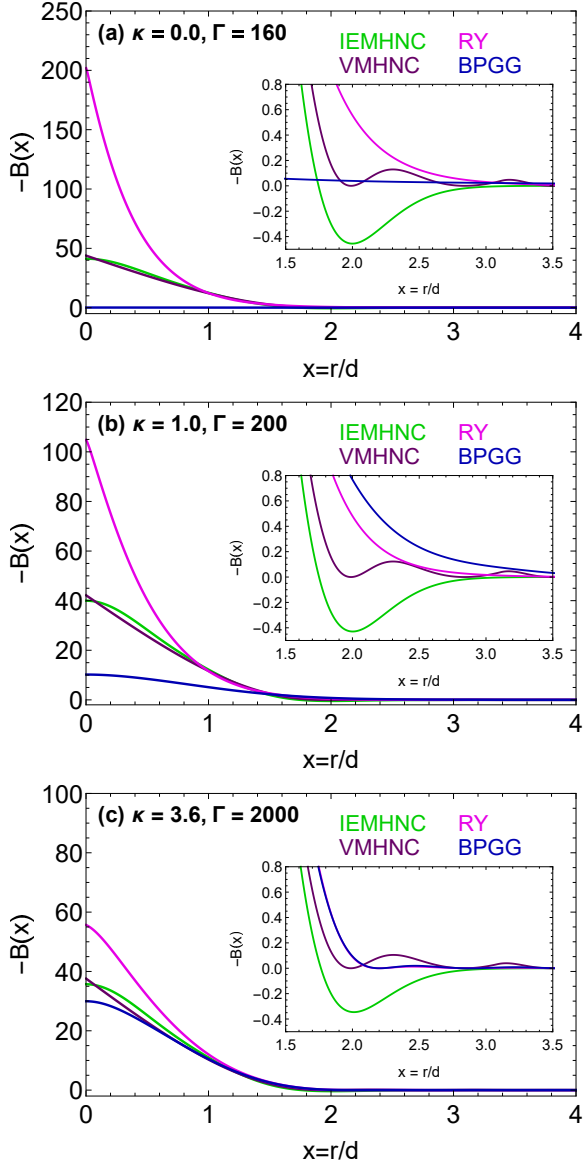}
\caption{Comparison between the IEMHNC (green), the VMHNC (purple), the RY (magenta) and the BPGG (blue) bridge functions, see Eqs.(\ref{eq:theory_iemhnc_bf},\ref{eq:theory_vmhnc_bfhspy},\ref{eq:theory_ry_bf},\ref{eq:theory_bpgg_bf}) respectively.
Results for state points close to the melting line ($\Gamma/\Gamma_{\mathrm{m}}>0.8$) with three different screening parameters: $\kappa=0$ (a), $\kappa=1$ (b), $\kappa=3.6$ (c). The insets reveal the intermediate and long range bridge function behavior by zooming-in on the $B(r)\sim0$ region.}\label{fig:discussion_bridge_function_comparison}
\end{figure}

\noindent A detailed comparison of different integral equation theory approaches at the level of bridge functions can shed ample light on the physical origin of their deviations from exact simulation results and on possible remedies of their deficiencies. Such comparison will be carried out for the IEMHNC, VMHNC, RY and BPGG bridge functions. It is illustrated in Fig.\ref{fig:discussion_bridge_function_comparison} for three state points close to melting at different screening parameters $\kappa=\{0.0,1.0,3.6\}$.

Given the high accuracy of the IEMHNC and VMHNC approaches at the level of the radial distribution function, it can be expected that the IEMHNC and VMHNC bridge functions are reasonably close to the unknown exact bridge function. From Fig.\ref{fig:discussion_bridge_function_comparison}, it becomes evident that, regardless of the YOCP state point, the two bridge functions are very close with the largest relative deviations concentrated in the oscillatory region where $B(r)\sim0$. \textbf{(i)} Exactly at the origin, both bridge functions acquire similar values consistent with Rosenfeld's zero-separation bridge function freezing criterion\,\cite{rosenfeld1981}. However, the small separation limit is approached in a different manner; almost quadratic with a zero slope at the origin within the IEMHNC approach versus almost linear with a non-zero slope at the origin within the VMHNC approach. These functional behaviors can be directly inferred from the Iyetomi construction\,\cite{iyetomi1992} and the Wertheim-Thiele solution\,\cite{wertheim1963,thiele1963}, respectively. \textbf{(ii)} The largest relative deviations are observed near the edge of the first co-ordination cell, where the IEMHNC bridge function becomes positive while the VMHNC bridge function remains negative. The existence of a narrow positive range in the IEMHNC bridge function for moderate to strong coupling can be proven by finding the roots of the polynomial pre-factor of the generating OCP bridge function, see Eq.(\ref{eq:theory_iemhnc_bOCP}). The non-positivity of the VMHNC bridge function regardless of the effective packing fraction can easily be proven by applying the inequality $\ln{y}\leq{y}-1$ in both branches of the generating HSPY bridge function, see Eq.(\ref{eq:theory_vmhnc_bfhspy}). Sign switching has emerged as a rather standard feature of extracted bridge functions for dense HS\,\cite{francova2011}, IPL\,\cite{restrepo1994} and Lennard-Jones\,\cite{restrepo1992} liquids that cannot be captured in the VMHNC approach. However, it is possible that both the extend and magnitude of this positive region is strongly overestimated within the IEMHNC approach owing to the relatively large bin width utilized by Iyetomi and collaborators\,\cite{iyetomi1992} in the determination of the OCP radial distribution function that was employed to infer the simulation extracted OCP bridge function. \textbf{(iii)} Small relative deviations persist in the long range, where the VMHNC bridge function possesses a decaying oscillatory behavior that is consistent with the exact asymptotic limit $B(r)\sim-(1/2)h^2(r)$, whereas the IEMHNC bridge function sharply decays to zero as artificially imposed in the OCP bridge function due to insufficient accuracy and to avoid over-complicated parameterizations\,\cite{iyetomi1992}.

On the other hand, the RY and BPGG approximations generate radically different estimates for the bridge function and appear to have opposite tendencies; the RY approach consistently overestimates the exact bridge function, whereas the BPGG approach systematically underestimates it. This underestimation becomes particularly severe in the OCP limit where the BPGG approach predicts $B(r)\approx0$. This result is consistent with the observation that the BPGG and HNC approximations are practically equivalent for $\kappa=0$. As the screening parameter increases, the differences between the RY and BPGG estimates for the bridge function are reduced in the hard-core region and practically eliminated for $r\geq 1.5d$. The similarity between the bridge function estimates of all four integral equation theory approaches for $r\geq{d}$ and $\kappa=3.6$ explains why they predict thermodynamic properties with comparable accuracy for $\kappa\geq3.0$.

\subsection{Deficiencies and further improvements of the IEMHNC bridge function}

\noindent Overall, the IEMHNC and the VMHNC bridge functions are very similar in the short range and exhibit appreciable deviations in the intermediate and long range. Undoubtedly, the IEMHNC approach and the VMHNC approach owe their great success to the very accurate description of the bridge function in the short range. Their deficiencies in the description of the intermediate and long range are reflected on their radial distribution functions, where the correlation void \& first maximum are predicted more accurately than the first minimum \& second maximum. In fact, it can be expected that the underlying isomorph invariance ansatz and quasi-universality conjecture both breakdown beyond the short range, as suggested by the prompt decay of the direct correlation function towards its potential-specific and state-point-specific asymptotic value, $c(r)\to-\beta{u}(r)$.

Therefore, theoretical efforts for further improvements of the IEMHNC bridge function should focus on the intermediate and long ranges. As discussed above and also reflected on the finite IEMHNC errors in the OCP limit, there are deficiencies in the OCP bridge function input. Consequently, we shall restrict ourselves to briefly presenting promising techniques and to demonstrating the efficiency of the simplest remedy. \emph{(a) A continuous interpolation between the IEMHNC approach and the VMHNC approach}. As inferred in sec.\ref{sec:discussion_bridge function}, the IEMHNC and the VMHNC bridge functions are expected to bracket the exact bridge function in the intermediate range. Given their near-identical short range behavior, a RY-type interpolation with the aid of a switching function that contains an adjustable parameter (as usual fixed by imposing thermodynamic consistency) should be beneficial. \emph{(b) A variational optimization procedure for the IEMHNC approach.} As detailed in sec.\ref{sec:results_thermo_consistency}, there is a ample space for improvement concerning the performance of the IEMHNC approach in terms of thermodynamic consistency. The latter can be improved by selecting the IEMHNC-YOCP system as reference and by minimizing the corresponding free energy functional. Given the non-IPL nature of the YOCP, this variational procedure will lead to two optimization equations in the spirit of the RHNC approach. \emph{(c) A cross-over modification of the IEMHNC approach}. Successful application of the SMSA approach to dense YOCP systems\,\cite{tolias2014,tolias2015} suggests that $c(r)=-\beta{u}(r)$ is accurate down to intermediate ranges. In view of the exact nonlinear closure equation, the asymptotic condition translates to the closure
\begin{equation}
B_{\mathrm{SMSA}}(r)=\ln{[g(r)]}-g(r)+1\,.\label{eq:cross_over_SMSA}
\end{equation}
A cross-over function can be employed to continuously interpolate between the IEMHNC closure and the asymptotic closure. At its simplest form, the cross-over function does not contain any adjustable parameters\,\cite{foiles1984,reatto1987}.

The proposed techniques (a,b) will manifoldly increase the computational cost of the IEMHNC approach, while they can only provide marginal improvements. In what follows, we shall apply the cross-over technique (c) to the IEMHNC approach. The adopted cross-over function $l(r)$ changes rapidly from zero to unity for distances beyond the first maximum position $r_{\mathrm{M}}$ of the radial distribution function. It reads as\,\cite{reatto1987}
\begin{equation}
l(r)=
\begin{cases}
0, & r\leq r^- \\
\displaystyle\frac{1}{2}\left[1+\tanh\displaystyle\left(\frac{r-R}{\sqrt{w^2-(r-R)^2}}\right)\right], & r^-<r<r^+ \\
1 , & r\geq r^+ \\
\end{cases}
\label{eq:cross_over_lr}
\end{equation}
with $r^\pm=R\pm w$, $R>r_{\mathrm{M}}$ the first post-maximum location satisfying $g(r)=1$ and $w=(3/2)(R-r_{\mathrm{M}})$. The cross-over IEMHNC closure becomes
\begin{equation}
B(r)=\left[1-l(r)\right]B_{\mathrm{IEMHNC}}(r)+l(r)B_{\mathrm{SMSA}}(r)\,,
\end{equation}
The cross-over IEMHNC algorithm requires an additional iteration circle that is necessary for the determination of the parameter $R$ starting from the bare IEMHNC estimate. The additional iterations have a negligible impact on the computational performances since they do not involve any integration or root-finding procedure and since two or three iterations are sufficient for convergence. Application of the cross-over technique to the VMHNC approach will modify both the closure and the optimization equation\,\cite{foiles1984}, which now read as
\begin{align}
&B(r)=\left[1-l(r)\right]B_{\mathrm{HSPY}}(r;\eta)+l(r)B_{\mathrm{SMSA}}(r)\,,\nonumber\\
&\frac{d\delta_\phi}{d\eta}-\frac{n}{2}\int\left[1-l(r)\right]\left[g(r)-g_{\mathrm{HSPY}}(r;\eta)\right]\frac{\partial{B}_{\mathrm{HSPY}}(r;\eta)}{\partial\eta}d^3r=0\,.\nonumber
\end{align}
As expected, the cross-over version of the IEMHNC approach leads to appreciably more accurate predictions concerning the position \& magnitude of the first minimum as well as the second maximum. Simultaneously, it slightly worsens the predictions concerning the position \& magnitude of the first maximum. In spite of this trade-off, there is an overall improvement manifested in the more accurate calculation of thermodynamic properties. The same does not apply for the cross-over VMHNC approach whose performance becomes slightly poorer.

The above investigation illustrates the numerous possibilities for even further improvement. A more detailed investigation is postponed until a more suitable parameterization is available for the OCP bridge function.

\section{Summary and future work}

\noindent Four advanced integral equation theory approaches have been applied to dense YOCP liquids and extensively compared: the recently-proposed IEMHNC approach which is based on the ansatz of exact bridge function isomorph invariance, the first-principle VMHNC approach which is constructed upon the conjecture of bridge function quasi-universality, the mixed RY approach which continuously interpolates between the HNC and PY bridge functions in a manner that ensures thermodynamic consistency, the mixed BPGG approach which introduces an adjustable parameter to the MS bridge function also in a manner that enforces consistency.

An exhaustive comparison with computer simulations at the level of radial distribution functions and of internal excess energies demonstrated that the IEMHNC \& VMHNC approaches possess comparable accuracies and are noticeably superior to the RY \& BPGG approaches across the dense liquid YOCP phase diagram. Moreover, the IEMHNC \& VMHNC approaches were concluded to possess a high degree of thermodynamic consistency between the virial and statistical routes to the isothermal compressibility without imposing it. In addition, extensive comparisons in terms of computational cost demonstrated that the IEMHNC algorithm converges $10-80$ times faster than the VMHNC algorithm. Furthermore, detailed comparison at the level of bridge functions revealed that the VMHNC is implicitly isomorph invariant and suggested that the IEMHNC \& VMHNC bridge functions are near-exact at the short range but suffer from minor deficiencies at the intermediate and long ranges. The latter conclusion led to the proposal of several optimization schemes and the numerical investigation of the cross-over IEMHNC approach, which leads to minor improvements without being accompanied by strong increase of the computational cost.

Future work will exclusively focus on the cumbersome task of extracting YOCP and OCP bridge functions from computer simulations with two main objectives in mind. (1) The extraction of exact YOCP bridge functions along phase diagram lines of constant excess entropy will confirm the isomorph invariance conjecture and quantify its range as well as its degree of validity. (2) The extraction of OCP bridge functions at multiple state points will enable a more accurate OCP parameterization that should further increase the accuracy of the IEMHNC approach.

\section*{Acknowledgments}

\noindent The authors would like to thank Prof. Jeppe Dyre for reading the manuscript and Prof. William Smith for suggesting the analytical HSPY extrapolation. The authors would like to acknowledge the financial support of the Swedish National Space Agency under grant no.\,143/16. Molecular dynamics simulations were carried out on resources provided by the Swedish National Infrastructure for Computing (SNIC) at the NSC (Link{\"o}ping University) that is partially funded by the Swedish Research Council through grant agreement no.\,2016-07213.

\bibliography{vmhnc_bib}

\end{document}